\documentclass[aps,prl,twocolumn,nofootinbib,showpacs,floatfix]{revtex4}
\usepackage{amsmath}
\usepackage{amssymb}
\usepackage{diagbox}
\usepackage{graphicx}
\usepackage{dcolumn}
\usepackage{bm}
\usepackage{xcolor}
\usepackage{verbatim}

\begin{document}


\title{Statistical mechanics of DNA-nanotube adsorption}

\author{Shushanik Tonoyan}
\author{Davit Khechoyan}
\author{Yevgeni Mamasakhlisov}
\affiliation{Department of Molecular Physics, Yerevan State University, A. Manougian Str. 1, 375025, Yerevan, Armenia}
\author{Artem Badasyan}
\email[Corresponding author's email:]{ abadasyan@gmail.com}
\affiliation{Materials Research Laboratory, University of Nova Gorica, \\ Vipavska 13, SI-5000 Nova Gorica, Slovenia}

\date{\today}

\begin{abstract}
Attraction between the polycyclic aromatic surface elements of carbon nanotubes (CNT) and the aromatic nucleotides of deoxyribonucleic acid (DNA) leads to reversible adsorption (physisorption) between the two, a phenomenon related to hybridization. We propose a Hamiltonian formulation for the zipper model that accounts for the DNA-CNT interactions and allows for the processing of experimental data, which has awaited an available theory for a decade.
\end{abstract}

\pacs{Valid PACS appear here}
\maketitle

Deoxyribonucleic acid (a.k.a. DNA) is a biomolecule, comprised of two 
polymer chains, stabilized by hydrogen bonds (H-bonds) in the 
perpendicular direction to its axis. If the H-bonds are broken and the 
two strands are separated, each single strand DNA (ssDNA) will remain 
stabilized by the $\pi$-stacking of neighbour nucleotides in the 
direction parallel to the axis. Polymer Physics, as a rule, considers 
linear polymers as one-dimensional objects in the absence of any of long-range interactions (including loops) \cite{cantor,gros,rubin}.
Another constituent of this complex under study, carbon nanotubes 
(CNT), are a system with cylindrical symmetry, that have unique 
electronic properties due to the relevant size-quantization effects, as 
well as outstanding mechanical properties thanks to their amazing 
structure \cite{saito,iijima}. Not surprisingly, CNTs have found 
numerous applications in varied areas such as nanoelectronics, 
medicine, environmental safety, and microbiology. Due to the large 
longitudinal to lateral dimension ratio, CNTs can be considered as 
one-dimensional objects as well. Attraction between these two rigid 1D 
objects results in the formation of a ssDNA-CNT complex, which, at a 
later stage of hybridization, serves as a landing site for free ssDNAs 
from solution. Once hybridized on the surface of CNT, 
double-stranded (ds) DNA undergoes a B to Z conformational transition that modulates the
dielectric environment of the single-walled CNT and allows for the 
optical detection of such event \cite{heller,maji}. The presence of an 
attracting 1D surface significantly enriches the phase 
diagram of adsorbed dsDNA \cite{kapri} and thus opens doors for 
numerous applications.

There are several reasons motivating the study the DNA-CNT complex. One 
is the insolubility or extremely poor solubility of CNTs, which imposes 
a considerable challenge when it comes to applications. Different 
techniques were developed to improve CNT dispersion including the use 
of surfactants, oligomers, biomolecules, polymer-wrapping, and chemical 
functionalization. One of the most efficient dispersing agents for 
water solutions is single-stranded DNA (ssDNA), which forms a (very) 
stable complex with CNTs \cite{vogel}. Another line of reasoning 
originates from the wide range of existing applications for the DNA-CNT 
complexes in various nanotechnologies. Short 
single-stranded DNA oligomers comprised of $\simeq 10$ nucleotides (nt) have been reported 
to be of exceptional relevance for many applications \cite{rox}.

Despite the fact that there are several reviews on biological 
\cite{cui} or biosensing \cite{zhu} applications for carbon 
nanomaterials, there is a negligibly small number of both theoretical 
and experimental studies devoted to the equilibrium picture of 
reversible adsorption (physisorption) of short single-stranded DNA 
oligomers on CNTs.

The standard approach in the field consists of the application of First 
Principles Calculations (mostly, using DFT software) to estimate the 
energies of interaction between the nucleotides and carbon-based 
substrates with and without water (see \emph{e.g.}, \cite{gao} and 
references therein). Another wide group of approaches is through the 
use of all-atom Molecular Dynamics simulations to reach conclusions 
about the thermodynamics of ssDNA-SWCNT interactions (see, e.g., 
\cite{pramanik},\cite{johnson}).

Recently, several phenomenological models have been employed towards 
the problem, mainly through the modifications of adsorption theories 
known from the past. Thus, to process the experimental data, a recent 
experimental study \cite{brunecker} has treated the adsorption of ssDNA 
oligomers and dimers as a simple chemical reaction. 

Kato \emph{et al} \cite{kato} have applied the Hill formula to estimate 
the adsorption free energy of single-stranded cytosine oligo-DNAs on 
single wall nanotubes (SWNT). In a recently published article 
\cite{butyrskaya}, an extended version of Langmuir’s approach is 
developed to describe the histidine and alanine adsorption on CNT. 
While simple and seemingly effective, adsorption isotherm models 
adopted to the biopolymer-CNT story suffer from the apparent and 
long-known limitations of the Hill-Langmuir approach in describing the 
cooperative adsorption of polymers. In particular, 
assumed presence of only two states (adsorbed and desorbed) is not 
justified. Instead, there are several minima 
present on the theoretical free energy landscape of short ssDNA oligomer adsorbed 
on CNT at room temperature \cite{johnson}, in agreement with 
experimental studies \cite{hughes}.

A general problem in the field is the absence of a Statistical 
Mechanical approach with a model Hamiltonian, that would provide 
a thermodynamic picture of reversible adsorption of a short 
ssDNA oligomer on CNT.

A recent experimental study serves as a bright example illustrating 
this apparent gap in knowledge. In 2009 Albertorio \emph{et al} 
reported an experiment on the association of ssDNA oligomers with CNTs  
\cite{albertorio}. The authors managed to process the results of 
kinetics experiments and to extract the association enthalpies with the 
help of the Eyring equation \cite{albertorio}. At the same time, they 
failed in extracting the data from the equilibrium measured curves of 
the temperature dependence of DNA/CNT fraction, because of the absence 
of a corresponding theory. The best they could do was to fit what they 
called sigmoidal function to their measured points, without providing 
any reasoning for the particular choice of function. Up to now, a 
theory that would provide a fit for experimental data like the one from 
Albertorio \emph{et al} \cite{albertorio} to a physical model with a 
well-defined microscopic Hamiltonian, has not been suggested.

In this Letter we describe the CNT-ssDNA physisorption phenomenon using 
the spin Hamiltonian formulation of a zipper model \cite{kittel} and 
validate theoretical results against experimental data. 
The zipper model is the limiting case of the Zimm-Bragg 
model, where the length of the chain is so short that there can be no 
more than one ordered and no more than one disordered region and, 
consequently will be no loops in the chain.

\begin{figure}
\includegraphics[width=0.48\textwidth]{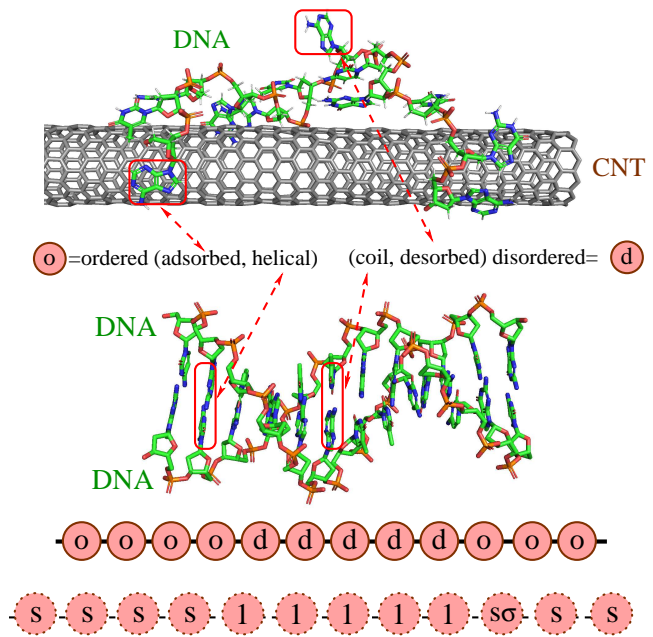}
\caption{\label{scheme} Scheme illustrating the similarity between the 
mixture of adsorbed and desorbed nucleotides of ssDNA on a CNT (above) and the helical and coil nucleotide pairs in dsDNA (below). Both systems can be reduced to the same sequence of ordered (o) and disordered (d) repeat units, giving rise to the sequence of statistical weights of Zimm-Bragg type.}
\end{figure}

We start by invoking the conceptual similarity between ssDNA adsorption on CNTs and the helix-to-coil transition or DNA melting (Fig.~\ref{scheme}). Indeed:
i) Due to the correlation of nucleotide conformations, ssDNA adsorption is promoted on the scale of the persistence length, and DNA melting is correlated (cooperative) on some spatial scale as well.
ii) The entropy of the adsorbed conformation is substantially smaller 
as compared to desorbed one because of the very different number of available 
conformations. In a similar way, helical repeat units are in a 
low-entropy conformation, as opposed to molten DNA.

iii) Short-range interactions (H-bonding between the strands of DNA and stacking between DNA and a nanotube) stabilize the association between the two one-dimensional systems.

Using the abovementioned similarity, we make use of models suggested in the past \cite{physa,PRL,europhys} and describe the adsorption of DNA on CNT with an energy function (Hamiltonian) that depends on the coarse-grained variables of the system. We do so by adopting the Potts-like spin model \cite{biopoly2,pre10,europhys} to the problem of DNA-CNT association. Employing the nearest-neighbor approximation we start with a spin Hamiltonian formulation, equivalent to the Zimm and Bragg model \cite{pre10}: 

\begin{equation} \label{ZBHam}
\begin{gathered}
H_{\text{ZB}}(\{\gamma_{i}\})= -U\sum\limits_{i=1}^{N}\delta(\gamma_{i},1)\delta(\gamma_{i+1},1) \\
\equiv -U\sum\limits_{i=1}^{N}\delta_i^{(2)},
\end{gathered}
\end{equation}
\noindent where $\gamma_{k}=1,2,...Q$ are spin variables describing the conformations of each of $i=1,2,...N$ repeat units (nucleotides), spin value $\gamma=1$ corresponds to the ordered (bound) conformation, other $Q-1$ values describe disordered (free) conformations; $U(>0)$ is the stacking energy per nucleotide. The Hamiltonian Eq.~\ref{ZBHam} leads (see \cite{pre10}) to a transfer-matrix with the characteristic equation
\begin{equation} \label{GMPC2chareq}
\Lambda^{2}-\Lambda(W-1+Q)+(W-1)(Q-1)=0,
\end{equation}
\noindent where $W=e^{U/T}$ and $T$ is temperature. Using mapping
\begin{equation} \label{mapping}
\frac{\Lambda}{Q}\rightarrow \lambda; \quad \frac{W-1}{Q}\rightarrow s; \quad \frac{1}{Q}\rightarrow \sigma,
\end{equation}
\noindent allows us to transform Equation~\ref{GMPC2chareq} into the original characteristic equation of ZB:
\begin{equation} \label{ZBchareq}
\lambda^{2}-\lambda(s+1)+s(1-\sigma)=0,
\end{equation}
\noindent with obvious roots
\begin{equation} \label{l12}
\begin{gathered}
\lambda_{1,2}(\sigma,s)=\frac{1}{2} \bigg[ 1+s \pm \sqrt{(1-s)^2+4\sigma s} \bigg]=\\
\frac{1}{2} \bigg[ 1+s \pm (1-s)\sqrt{1+\frac{4\sigma s}{(1-s)^2}} \bigg].
\end{gathered}
\end{equation}
\noindent Since the Thermodynamics is fully determined by the characteristic equation of the model, Eq.~\eqref{ZBHam} can be considered the Hamiltonian of the ZB model \cite{pre10}. The solutions of Eq.~\eqref{ZBchareq} are eigenvalues that provide the link between model parameters $s,\sigma$ and the partition function:
\begin{equation} \label{partitionfunction}
\begin{gathered}
Z(\sigma,s)=c_1\lambda_{1}^{N}+c_2\lambda_{2}^{N}= 
\lambda_1^N \Big[ c_1+c_2 e^{-N/ \xi} \Big],
\end{gathered}
\end{equation}
\noindent where $ N $ is the number of repeat units, $c_1=\frac{1-\lambda_2}{\lambda_1-\lambda_2}$, $c_2=\frac{\lambda_1-1}{\lambda_1-\lambda_2}$ (\cite{polsher}) and  
\begin{equation} \label{xizb}
\xi(\sigma,s)=1/\log(\lambda_1/\lambda_2)
\end{equation}
is the spatial correlation (or persistence) length, a curve with its maximum at the transition point. For finite correlation lengths ($\xi<\infty$) the effect of the second eigenvalue on the partition function decreases exponentially with the increase of $N$:
\begin{equation} \label{pf-large}
\begin{gathered}
Z(\sigma,s) \xrightarrow[N\gg\xi]{} c_1 \lambda_{1}^{N} \approx \lambda_{1}^{N}.
\end{gathered}
\end{equation}
\noindent This is the regular, large $N$, limit of the Zimm-Bragg theory, meaningful for longer polymer chains, but not applicable to our problem of interest: oligomer DNA adsorption on carbon nanotubes. In their experiment, Albertorio \emph{et al} used DNA oligomers of 12 nucleotide bases long, which is on the order of the Kuhn length of a single strand DNA (ssDNA), \emph{i.e.} $N \sim 2\xi$. Therefore we need to return to Eq.~\eqref{l12} and apply the single-sequence approximation of the Zimm-Bragg model. At the heart of the single-sequence approximation is the impossibility of having more than one uninterrupted sequence of helical (ordered) repeat units due to small system sizes ($N<2\xi$). For this regime, the role of the small parameter is played by 
\begin{equation} \label{smallp}
\frac{4\sigma s}{(1-s)^2}\ll 1.
\end{equation}
\noindent After resolving Eq.~\eqref{l12} into the Taylor series by this small parameter and keeping the first terms, we obtain the eigenvalues
\begin{equation} \label{l12limit}
\lambda_{1}(\sigma,s)=1+\frac{\sigma s}{1-s} \quad ;  \quad \lambda_{2}(\sigma,s)=s-\frac{\sigma s}{1-s}.
\end{equation}
\noindent When inserted into Eq.~\eqref{partitionfunction}, we obtain:
\begin{equation} \label{pfzip0}
Z(\sigma,s)=\frac{(1-s+\frac{\sigma s}{1-s})(1+\frac{\sigma s}{1-s})^{N}+\frac{\sigma s}{1-s}s^{N}(1-\frac{\sigma}{1-s})^{N}}{1-s+\frac{2\sigma s}{1-s}}.
\end{equation}
\noindent After resolving the powers into series, rearranging the results and keeping only terms linear in $\sigma$, we obtain 
\begin{equation} \label{pfzip}
Z(\sigma,s)=1+\frac{\sigma s}{(1-s)^2}(N-1-Ns+s^N)+O(\sigma^2).
\end{equation}
The order parameter (helicity degree) is calculated from the partition function as
\begin{multline} \label{op}
\theta(\sigma,s)=\frac{1}{N}\frac{\partial \log Z(\sigma,s)}{\partial \log s}=\frac{s}{NZ(\sigma,s)}\frac{\partial Z(\sigma,s)}{\partial s}=\\
\frac{\sigma s}{N(s-1)^3}\bigg[ \frac{(N-1)(s^{N+1}-1)-s(N+1)(s^{N-1}-1)}{1+\frac{\sigma s}{(s-1)^2}(N-1-Ns+s^N)} \bigg].
\end{multline}
Eq.~\eqref{op} is a well-known helicity degree formula for a zipper 
model, appearing in many papers and books \cite{polsher,qian}. Thus, 
using the analogy between the adsorption of one DNA strand onto another 
in double-stranded DNA and single-strand DNA adsorption onto a 
nanotube, we have derived Eq.~\eqref{op} as a theoretical formula, 
describing the order parameter, the fraction of adsorbed nucleotides. 
The expression contains oligomer length (in nucleotides) as a parameter, since we have 
explicitly taken into account finite-size effects that dominate the 
behavior of zipper model. Before the application of Eq.~\eqref{op} to 
data treatment, we need to translate the Zimm and Bragg parameters $s$ and 
$\sigma$ into experimental variables. There have been 
several definitions of these parameters in the past \cite{polsher}. We 
stick to one the most general definitions from 
Ref.~\cite{polsher} and following our previous publications \cite{biopoly2,physa,pre10}, consider the stability parameter $s$ as 
a statistical weight in terms of a (Gibbs or Helmholtz) free energy change between the bound and unbound states, as:
\begin{equation}
s=\exp\left(-\frac{\Delta G}{RT}\right)=\exp\left(-\frac{\Delta 
H-T\Delta S}{RT}\right),
\end{equation} 
\noindent where the enthalpy of binding per nucleotide and the entropic price of 
adsorption per nucleotide can be expressed through $U$ and $Q$ as
\begin{equation} \label{hs}
\Delta H=-U \quad \textrm{and} \quad \Delta 
S=-R\ln Q,
\end{equation}
\noindent respectively \cite{biopoly2,physa,pre10}.
The cooperativity parameter $\sigma$, by its definition, describes how much the original probability of bounded region growth, $s$, is hindered by the fact that there is no preceding bounded repeated unit. It can be estimated (see \cite{biopoly2,physa,pre10}) as 
\begin{equation} \label{sgma}
\sigma=Q^{1-l},
\end{equation}
\noindent where $l$ (=6 nucleotides) is the persistence 
length of ssDNA. \footnote{Persistence length depends 
on sequence, with typical range of values between 3 to 10 nt. We chose 
the $l=6$ nt value to simplify the expressions.} After inserting the definitions of $s$ and $\sigma$ into Eq.~\eqref{op}, we arrive at 
\begin{equation} \label{opfit}
\theta(s,\sigma)=\theta(T,U,Q,l=6,N=12)=\theta(T,U,Q),
\end{equation}
\noindent a formula, that contains only two free parameters: $U$ and $Q$.

\begin{figure}
\includegraphics[width=0.425\textwidth]{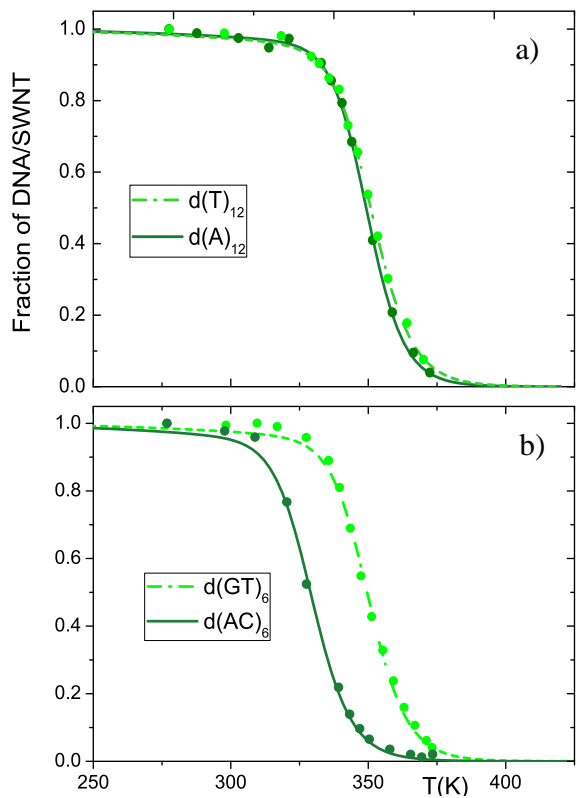}
\caption{\label{fit1} The fraction of total initial DNA/SWNTs as a function of temperature for a) poly$d(A)_{12}$, poly$d(T)_{12}$ and b) poly$d(AC)_{6}$, poly$d(GT)_{6}$, fitted by zipper model. Dots are experimental points from \cite{albertorio}, lines correspond to fitted Eq.~\eqref{opfit}. Two curves (poly $d(C)_{12}$ and poly $d(G)_{12}$) are not shown due to the low quality of experimental data, which resulted in large fitting errors.}
\end{figure}

\begin{table}  
\begin{center}
\begin{tabular}{ |c|c|c||c|c|c| } 
 \hline
		& $U$ & $Q$ & $\Delta H$ & $\Delta S/R$ & 
		$\sigma$ \\ 
 \hline
 $d(A)_{12}$ & 3.28(0.06) & 29(1.7) & -3.28 & -3.37 & $4.8\mathrm{e}{-8}$\\ 
 \hline
 $d(T)_{12}$ & 3.03(0.06) & 22(1.3) & -3.03 & -3.09 & $1.9\mathrm{e}{-7}$\\
 \hline 
 $d(AC)_{6}$ & 2.77(0.10) & 21(2.1) & -2.77 & -3.05 & $2.5\mathrm{e}{-7}$\\ 
 \hline
 $d(GT)_{6}$ & 3.13(0.09) & 25(2.4) & -3.13 & -3.22 & $1.0\mathrm{e}{-7}$\\ 
 \hline
\end{tabular}
\end{center}
\caption{
\label{tabfit} 
Parameter values resulting from fit. All quantities are per mole and 
per base of nt; $U$ and $\Delta H$ 
in the units of \emph{kcal$\times$mole$^{-1}$$\times$base$^{-1}$}; fit error shown in brackets. First two 
columns result from the fit of Eq.~\eqref{opfit}, other three columns 
recalculated with Eqs.~\eqref{hs},\eqref{sgma}.
Heteropolymers have been fitted, assuming the length of 12 nucleotides, 
and therefore the fitted parameters correspond to averaged quantities. 
}
\end{table}

In order to check how adequately the proposed theory describes the 
phenomenon, we have chosen an experimental study which reports the 
measured fraction of adsorbed nucleotides, namely, the study by 
Albertorio \emph{et al} \cite{albertorio}. In their study, a solution 
of 12-base-long \footnote{Many authors have reported the same 
choice of oligomer lengths about 10-12 nt to be optimal for 
applications. This is the very scale of Kuhn length for ssDNA and can 
serve as a possible explanation of such choice, since sequence 
specificity, recognition and sensing is optimal at exactly this scale. 
Yet another property of ssDNA oligomers is the absence of loops below 
12 nt \cite{johnson}, which is logical: system is too rigid to wrap 
around CNT.}
single stranded DNA homopolymers consisting of poly $d(A)_{12}$, poly $d(T)_{12}$, poly $d(C)_{12}$, and poly $d(G)_{12}$, as well as regular heteropolymers poly $d(AC)_{6}$ and poly $d(GT)_{6}$ was added to single-wall carbon nanotubes (SWNT) at a 1:1 DNA:SWNT mass ratio. The DNA/SWNT mixture was sonicated and then the bundles of non-dispersed nanotubes and the remaining free DNA were removed. The thermal stability of the obtained hybrids was quantified indirectly by measuring the extent to which 12-base-long ssDNA polymers dissociated from the nanotubes after incubation in an aqueous buffer solution at different temperatures in the 4-99$^o$C range for 10 min by the detection of optical absorption at 815 nm. 

We have digitized Figures 2 and 3 of Ref.~\cite{albertorio} reporting 
the temperature-dependent fractions remaining in solution and fit them 
with Eq.~\eqref{opfit}. Results of the fit are shown in 
Figure~\ref{fit1} and in Table~\ref{tabfit}. As one can see, the fit is 
close to perfect, which, considering that there are just two free 
parameters, ensures the validity of the statistical approach developed. 
The values of fitted energies (enthalpies) of adsorption 
(Table~\ref{tabfit}) are all about -3 $kcal/mol$ per 
nucleotide, in agreement with previously reported 
values \cite{johnson,albertorio}. The obtained 
adsorption parameters are lower than the same parameters for the dsDNA 
melting. For example, $\Delta H_{ads}\approx-3.28~kcal/mol$ against 
$\Delta H_{melt}\approx-8~kcal/mol$ and $\Delta 
S_{ads}/R\approx -3.37$ against $\Delta S_{melt}/R\approx -11$ for the $AT$ 
base pair. The halved enthalpy can be explained by the fact that 
the ssDNA adsorbed on the surface of CNT is not stabilized by H-bonds, 
which are known to contribute roughly 50$\%$ to free energy 
\cite{cantor}. The absence of H-bonds has also entropic consequences: 
ssDNA adsorbed on CNT has higher freedom (number of available 
conformations), as compared to ssDNA adsorbed on a complementary ssDNA 
and fixed by H-bonds. Regarding the particular ordering of adsorption enthalpies by nucleotide type, there is a long history of contradictory reports, as is nicely reviewed by Pramanik and Maiti \cite{pramanik}. Based on the data provided in Ref.~\cite{albertorio}, we cannot support a particular view on the adsorption strength ordering of nucleotides, since the experimental curves for poly $d(C)_{12}$ and poly $d(G)_{12}$ span outside the experimentally accessible range of temperatures, and their desorption is incomplete (Figure 2 of \cite{albertorio}), thus essentially decreasing the quality of the fit (not shown). However, based on the available data on poly $d(A)_{12}$ and poly $d(T)_{12}$, our analysis confirms purines having larger enthalpy of adsorption as compared to pyrimidines (\emph{i.e.} $A > T$ order), in agreement with many reported studies (see Ref.~\cite{pramanik} and references therein). Since we are not aware of any other published data on the temperature dependence of the ratio of adsorbed nucleotides on CNT, more experimental data are needed to make conclusion about the order of adsorption strengths for different nucleotides. 

However, not only are the fitted numbers relevant \emph{per se}, but also the model itself, since it provides a language for the treatment of the phenomenon. For instance, in the same paper, Albertorio \emph{et al} \cite{albertorio} also mentions problems with the stability of adsorbed DNA because of desorption. They have introduced extra stabilization by increasing the free DNA concentration in solution. This stabilizing effect is reported, but not explained or modelled. Instead, a line of naive argumentation could lead to the opposite expectations that the presence of extra free ssDNAs in solution will result in the promotion of ssDNA-ssDNA interactions, which should introduce a destabilizing effect onto the ssDNA-CNT complex because of obvious competition between the two targets for adsorption. In view of our previous studies of the osmotic stress effects onto DNA conformations \cite{PRL}, the reported increase in stability of bound conformations finds its explanation as arising because of the increased osmotic stress due to the increased excluded volume effects (crowding) from the free DNA added. 
From the physical point of view, up to Eq.~\eqref{pf-large}, the model 
allows for both cooperative (for small nontrivial $\sigma$ at 
$s\approx1$) and phase transition (at $\sigma=0$ and $s=1$) pictures.  
But after we assume that chain (oligomer) sizes are of the order of Kuhn length 
and accept the single-sequence approximation through 
Eq.~\eqref{smallp}, the resulting zipper model 
describes the finite-size effects of the transition. Small, but non-zero values of fitted $\sigma$ in 
Table~\ref{tabfit} reflect the deviation from the ideal phase 
transition picture.
Thus, by providing a Statistical Mechanical Hamiltonian to describe the 
DNA-CNT interaction, which is at the heart of numerous 
Nano(Bio)technologies, we contribute towards a better understanding of the principles behind the relevant biotechnologies and suggest a route to the predictable design of nanodevices.

\begin{acknowledgments}
A.B. and D.K. acknowledge support from Erasmus+ project 
2018-1-SI01-KA107-046966; Y.M and Sh.T. acknowledge financial support from 
Enterprise Incubator Foundation (EIF) Research Faculty 2019 award; Sh.T. and D.K. acknowledge financial support from RA MES State Committee of Science through the research project 16YR-1F046. A.B. acknowledges financial support from the Slovenian Research Agency through project J1-1705.
\end{acknowledgments}

\end{document}